\documentclass[12pt,oneside,fleqn]{article}
\usepackage{epsf}
\usepackage{amsfonts, bm}
\usepackage{amsmath}
\usepackage{amssymb}
\usepackage[latin2]{inputenc}
\usepackage[T1]{fontenc}
\usepackage{epsfig}
\usepackage{mathrsfs}
\usepackage{verbatim}

\newcommand{\de}{\text{d}}

\newcommand{\lb}{\lambda}

\newlength{\dinwidth}
\newlength{\dinmargin}
\DeclareMathAlphabet{\scr}{U}{rsfs}{m}{n}

\begin{document}
\renewcommand\thefootnote{\fnsymbol{footnote}}
\begin{center}
{\large {\bf Remark on a group-theoretical formalism for quantum mechanics and the quantum-to-classical transition}}\\
\vspace{0.5cm}
J. K. Korbicz$^{1,2}$\footnote{jarek@itp.uni-hannover.de} and M. Lewenstein$^{2,1}$\\
\vspace{0.3cm}
$^1$ Institut f\"ur Theoretische Physik, Universit\"at Hannover, D-30167
Hannover, Germany\\

$^2$ ICREA and ICFO--Institut de Ci\`{e}ncies Fot\`{o}niques, Mediterranean Technology Park, 08860
Castelldefels (Barcelona), Spain
\end{center}
\setcounter{footnote}{0}
\renewcommand\thefootnote{\arabic{footnote}}

\abstract{We sketch a group-theoretical framework, based on the Heisenberg-Weyl group, encompassing both quantum and classical statistical descriptions of unconstrained, non-relativistic mechanical systems. We redefine in group-theoretical terms a kinematical arena and a space of statistical states of a system, achieving a unified quantum-classical language and an elegant version of the quantum-to-classical transition. We briefly discuss the structure of observables and dynamics within our framework.}

\section{Introduction.}
Since the seminal works of Weyl \cite{Weyl} and Wigner \cite{Wigner}, the fundamental role of group theory in quantum mechanics has become an established fact. The aim of the present work is to remark that it is possible to redefine, using solely group-theoretical notions, the mathematical representations of kinematical arena and state-space in non-relativistic quantum mechanics, such that i) a unified language for quantum and classical statistical descriptions is obtained; ii) there is a natural transition mechanism, leading from a more generic quantum state-space to a classical one. The latter is a mathematically precise formulation of the classical limit of quantum theory at the level of statistical descriptions. It constitutes the main result of our paper. Since the literature on the subject is enormous, let us state it clearly that our work is neither meant to be an overview of the quantization methods, nor even an overview of various realisations of classical limits of quantum mechanics. For that see, e.g., Refs. \cite{Wernerchuj, Landsman} and the references therein. 

In classical physics, statistical description of mechanical systems is given in terms of probability measures $\mu$ on a phase space $\Gamma$. On the other hand, in quantum mechanics, statistical properties are encoded into density matrices $\varrho$ acting on a Hilbert space $\mathcal{H}$. A natural question arises how to connect these two such seemingly different formalisms through a sort of a quantum-to-classical transition. Obviously, such a transition must exists, as indirectly proven by numerous examples and more or less heuristical arguments within all possible approaches to quantum theory. The question is rather how to write it in a clear, precise manner in a hope to shed some light on {\it how} the classical world appears. One strategy, which we adopt in this work, is first to try to find a common theoretical framework for both quantum and classical theories and then search for a transition mechanism within this framework. The standard realization of such approach is to use Wigner \cite{WignerF} or Moyal functions \cite{Moyal}, associating with every density matrix $\varrho$  a phase-space pseudo-probability distribution $W_\varrho$ and a corresponding pseudo-characteristic function $\chi_\varrho$ respectively (e.g. Ref. \cite{Schleich} contains a modern exposition). However, it is a well known fact \cite{WignerF, Schleich} that both $W_\varrho$ and $\chi_\varrho$ fail to satisfy the positivity conditions that possess their classical analogs. Moreover, there seems to be no direct relation between positivity of Wigner or Moyal functions and ``classical behaviour'' of density matrices. Whether this is a drawback or not in the context of providing a unified framework for both quantum and classical statistics is perhaps a matter of taste, but let us assume that it is.  

The aim of our work is to point out that such a unified language with a clearly visible quantum-to-classical transition mechanism is achieved if instead of working with density matrices and probability measures one performs (generalized) Fourier transform and works with (generalized) characteristic functions. The structure, which emerges after such Fourier transform is {\it universal}---in {\it both} quantum and classical cases it consists of a certain group $G$, serving as a sort of ``kinematical arena'' (instead of a phase space $\Gamma$ or a Hilbert space $\mathcal{H}$) and a set of normalized, positive-definite functions $\phi$, representing statistical states. Recall \cite{Folland} that a complex function $\phi$ on a group $G$ with a Haar measure $\text{d}g$ is called positive-definite if it is bounded, continuous, and satisfies: \begin{equation}
\iint \text{d}g\text{d}h \overline{f(g)}\phi(g^{-1}h)f(h)\ge 0 \ \ \text{for any}\  f\in L^1(G); \label{PD}
\end{equation}  
by normalization we mean here that:
\begin{equation}
\phi(e)=1. \label{norm}
\end{equation}
In the case when $G=\mathbb{R}^n$, Eqs. (\ref{PD}) and (\ref{norm}) represent familiar properties of a characteristic function of a classical probability distribution $\mu$ on $\mathbb{R}^n$ \cite{Abram}. As we shall show in the sequel, when $G$ is non-Abelian, $\phi$ can still be viewed as a characteristic function, which we call ``non-commutative characteristic function'', but of a quantum probability distribution $\varrho$. And this is precisely the difference between classical and quantum statistics in the emerging formalism---in quantum case group $G$ is necessarily non-commutative, whereas in classical case it is Abelian.

In order to avoid the need for performing Fourier transform forth and back and thus achieve some, at least formal, conceptual simplicity, we propose to make one step further and {\it postulate} that a suitable group $G$ together with a (sub)set of normalized, positive-definite functions $\phi$ should be taken as a basis of statistical description of mechanical systems, {\it both} classical and quantum. Thus, the change of concepts we propose to examine is the following: i) as the kinematical arena of the statistics we consider a certain group $G$, called ``kinematical group'', together with its irreducible unitary representations; ii) as mathematical representatives of statistical states of the system we consider normalized, positive-definite functions on $G$. We show that such a unified formalism is indeed equivalent to the standard ones, upon a correct choice of $G$.

The kinematical group of our approach should not be confused with a group of symmetry transformations of neither kinematics nor dynamics of the theory (for an alternative programme for quantum theory, where $G$ is taken to be a group of dynamical symmetries see Ref. \cite{Mielnik}). Its role is rather to serve as a background for statistics (just like classical phase space $\Gamma$ or Hilbert space $\mathcal{H}$), encoding statistical properties of the system in a proper way. For example, as we show in this work, in the case of unconstrained mechanical systems (particles) with $n$ degrees of freedom the right kinematical group turns out to be the Heisenberg-Weyl group $H_n$ \cite{Weyl, Folland, Perelomov}. Intuitively, this can be understood in the following way: since in the standard formulation of quantum mechanics $H_n$ is a realization of Heisenberg uncertainty principles, it turns out that it is this information that is enough to re-produce quantum statistics. Hence, we invert the usual role of the Heisenberg-Weyl group and instead of treating it as a mere consequence of the uncertainty principles, we propose to look at it as the source of the latter. In this sense our approach bears some similarities to Klein's Erlangen programme---as we shall see in the sequel kinematical group determines convex geometry of statistical states. However, let us stress again that, unlike in Erlangen programme, our kinematical group is \emph{not} a group of symmetries.
  
When viewed from the perspective of the standard formalism, the resulting approach closely resembles Weyl quantization \cite{Weyl}. The difference is that instead of quantizing functions on the classical phase space $\Gamma=\mathbb{R}^{2n}$, which can be viewed as an Abelian group, we quantize functions on the non-Abelian group $H_n\simeq\mathbb{R}\times \Gamma$. This seemingly subtle difference produces, as we will show, some interesting changes. For example, the Hilbert space structure of quantum theory does not have to be fully and independently postulated, but to some extent follows naturally from the consideration of irreducible representations of $H_n$, if the latter is treated as a fundamental entry of the formalism. Second, and perhaps more importantly, one clearly sees how the classical statistical state-space naturally emerges from the quantum one as $H_n$ collapses to one of its Abelian subgroups. Thus, the essence of quantum-to-classical transition in our framework is the restoration of commutativity of the kinematical group. In the context of more modern versions of quantum theory, our formalism is closely related to the algebraic approach (see e.g. Ref. \cite{Haag} for a deep exposition and Ref. \cite{Landsman} for the latest trends) and, in fact, can be viewed as a concrete, but rather non-standard, realization of the latter.    

Finally, let us mention that in physical literature, the group-theoretical formalism that we develop in the present paper was in fact initiated by Gu in Ref. \cite{Gu}. Especially in the context of providing a more coherent, as compared to the standard Wigner and Moyal functions, way of describing both quantum and classical statistics. However, Gu did not fully perform the reformulation of the theory and concentrated mostly on practical problems, treating non-commutative characteristic functions rather as secondary objects with respect to the usual density matrices. Neither did he examine the representation of observables and the classical limit (on the level of kinematics) in the resulting formalism. In the present work we explicitly carry over the mentioned conceptual change and treat from the beginning non-commutative characteristic functions as primary objects of the theory, while density matrices or probability measures as secondary. 

The plan of the work is the following: in Section \ref{HW} we recall the basic properties of the Heisenberg-Weyl group. In Section \ref{GR} we develop the group-theoretical formalism and present our version of the quantum-to-classical transition. In Section \ref{Obs} we sketch the group-theoretical representation of observables and briefly comment on the dynamics in our scheme. Then, in Section \ref{Lim}, we show with two physical examples how the classical transition in our language works in practice. It should be kept in mind however that the objective of this work is not to develop practical methods of calculating classical limits of density matrices, but rather to put quantum-to-classical transition in a mathematically rigorous form. The concluding remarks are gathered in Section \ref{Concl}.

\section{Heisenberg-Weyl group}\label{HW}
The basic object of our study will be the Heisenberg-Weyl group $H_n$, where $n$ is the number of degrees of freedom of the considered mechanical system. Thus, at this moment we fix the kinematical group: $G=H_n$. The group $H_n$ can be identified with a space $\mathbb{R}\times\mathbb{R}^n\times\mathbb{R}^n$, equipped with the following multiplication law:
\begin{equation}
(s,\bm{\eta},\bm{\xi})\cdot (s',\bm{\eta}',\bm{\xi}'):=\Big(s+s'+\frac{1}{2}\omega[(\bm{\eta}, \bm{\xi}),(\bm{\eta}', \bm{\xi}')],\bm{\eta}+\bm{\eta}',\bm{\xi}+\bm{\xi}'\Big)\,,\label{prawo}
\end{equation}
where $(s,\bm{\eta},\bm{\xi})$ are the coordinates and:
\begin{equation}
\omega=\left(\begin{array}{cc} 0 & -{\bf 1}\\
                          {\bf 1} & 0 \end{array}\right)\,.
\end{equation}
In the sequel we will interchangeably denote group elements by $g,h,\dots$ or by the corresponding coordinates. The Haar measure $\text{d}g$ on $H_n$ is just $\text{d}s\, \text{d}^n \bm{\xi}\, \text{d}^n \bm{\eta}$. 

In what follows we will need irreducible, unitary (and strongly continuous) representations of $H_n$. They  are characterized by the Stone-von Neumann Theorem (see e.g. Refs \cite{Folland, Perelomov}). Let us briefly recall their structure. There is a family of infinite-dimensional representations $T^\lambda$ , $\mathbb{R}\ni\lb\ne 0$:
\begin{equation}\label{Tl}
T^\lambda(s,\bm{\eta},\bm{\xi})=\text{e}^{-\text{i}\lambda s} \,\text{exp}\Big[\frac{\text{i}}{\lambda}(\eta_j\hat{q_j}-\xi_j\hat{p_j})\Big]
\end{equation}
(throughout the work the repeated indices are summed over), where the self-adjoint generators $\hat{{\bf q}}, \hat{{\bf p}}$ satisfy on the common domain:
\begin{equation}\label{ccr}
[\hat{q_j},\hat{p_k}]=\text{i}\lb \delta_{jk},
\end{equation}
and a family of one-dimensional representations $T^0_{{\bf q},{\bf p}}$, labelled by points $({\bf q},{\bf p})\in\mathbb{R}^{2n}$:
\begin{equation}
T^0_{{\bf q},{\bf p}}(s,\bm{\eta},\bm{\xi}):=\text{e}^{\text{i}(\eta_j q_j-\xi_j p_j)}\,.\label{T0}
\end{equation}
From experiment we know that the representation realized in Nature is $T^{\hbar}$ with $\lambda=\hbar$.

\section{Group-theoretical approach}\label{GR}
In order to expose the quantum-to-classical transition mechanism, we adopt the strategy that first a suitable unified quantum-classical statistical framework should be developed. Traditionally, a search for such a framework has been interpreted as a search for quantum analogs of classical probability distributions. In case of mechanical systems one thus follows Moyal \cite{Moyal}, and defines a phase-space characteristic function (also known as the Moyal function) corresponding to a given density matrix $\varrho$ by:
\begin{equation}\label{chi}
\chi_\varrho (\bm{\eta},\bm{\xi}):=\text{tr}\Big (\varrho \,\text{exp}\Big[\frac{\text{i}}{\hbar}(\eta_j\hat{q_j}-\xi_j\hat{p_j})\Big]\Big).
\end{equation}
One passes then to its Fourier transform---the Wigner function \cite{WignerF, Moyal}:
\begin{equation}\label{Wigner}
W_\varrho ({\bf q},{\bf p}):=\int\frac{\de^n\bm{\xi}\de^n\bm{\eta}}{(2\pi\hbar)^{2n}}\text{e}^{-\frac{\text{i}}{\hbar}(\eta_j q_j-\xi_j p_j)}\chi_\varrho (\bm{\eta}, \bm{\xi}),
\end{equation}
in the hope to obtain an analog of a classical probability distribution. However, this attempt fails---as we have mentioned in the Introduction, Wigner function (\ref{Wigner}) is generically non-positive on the classical phase space $\Gamma=\mathbb{R}^{2n}$. Moreover, there seems to be no universal relation between positivity of $W_\varrho$ and ``classical behaviour'' of density matrices: there are density matrices showing what is generally accepted as ``genuine quantum behaviour'', and nevertheless possessing positive Wigner functions (for example, so called, squeezed states \cite{Schleich}). 

There has been developed some methods to get around the above difficulty. One of them is to replace the abstract definitions (\ref{chi}) and (\ref{Wigner}) by operational ones, i.e. involving prescribed interaction with an external reference particle. This allows one to construct a positive phase-space probability distribution (see e.g. Ref. \cite{Wodkiewicz}). Another way of producing a positive phase-space probability distribution is to use the Glauber-Sudarshan coherent states $|\alpha\rangle$ \cite{Sudarshan, Perelomov}, and assign to each density matrix $\varrho$ the Husimi function, also known as the $Q$-representation, $\langle \alpha |\varrho \,\alpha\rangle$ (see e.g. Ref. \cite{Davidovic}).

In contrast to the approaches mentioned above, we propose, following  Ref. \cite{Gu}, an alternative way towards the unification of languages of quantum and classical statistics. Instead of searching for positive phase-space probability distributions for density matrices, let us rather change the object of our interest and look at the properly generalized characteristic functions. Observe that the lack of positivity of $W_\varrho({\bf q},{\bf p})$ is mathematically related to the fact that characteristic function (\ref{chi}) generically is not positive-definite (c.f. Eq. (\ref{PD})) on $\mathbb{R}^{2n}$ \cite{modified}. Gu noted in Ref. \cite{Gu} that if one replaces the standard definition (\ref{chi}) by, in our opinion, more logical one:
\begin{equation}\label{fdt}
\phi_\varrho(g):=\text{tr}\big[\varrho T^{\hbar}(g)\big]=\text{e}^{-\text{i}\hbar s}\chi_\varrho (\bm{\eta},\bm{\xi})\, ,
\end{equation}
then such defined function $\phi$ is positive-definite on $H_n$ and is also normalized: $\phi(e)=1$, where $e=(0,\bm{0},\bm{0})$ is the neutral element. Hence, as we have mentioned earlier, it possesses all the features of a classical characteristic function of a probability distribution. However, unlike the latter, $\phi_\varrho$ is defined on the non-Abelian group $H_n$, rather than on $\mathbb{R}^{2n}$. This justifies the terms ``kinematical group'' for $H_n$ and ``non-commutative characteristic function'' for $\phi_\varrho$. Note that one cannot straightforwardly apply the usual Fourier transform to $\phi_\varrho$, as it was done in Eq. (\ref{Wigner}). Rather Eq. (\ref{fdt}) is a non-commutative Fourier transform of the density matrix $\varrho$. 

Motivated by the above observation, we propose to examine the following alternative construction of quantum statistics of an unconstrained mechanical system\footnote{For a similar approach, based on covariance systems see Ref. \cite{Naudts}}: 
\begin{itemize}
\item treat the group $H_n$ as the basic entry of the formalism, which sets up the kinematical arena; 
\item take as statistical states of the system (abstract at this moment) normalized, positive-definite functions $\phi$ on $H_n$; the set of such functions will be denoted by $\mathcal{P}_1(H_n)$ and it is a convex subset of the set of all continuous, bounded functions on $H_n$. 
\end{itemize}
Heuristically, the appearance of $H_n$ rather than $\Gamma$ may be explained as follows: $H_n=\mathbb{R}\times \mathbb{R}^{2n}\simeq\mathbb{R}\times \Gamma$, so we may view (to some extent) $H_n$ as an extension of the classical phase space. The additional degree of freedom, labelled by $s$, per point of $\Gamma$ can be then attributed to quantum-mechanical phase, which is supported by the multiplication law (\ref{prawo}) and the form of generic representations (\ref{Tl}). This phase degree of freedom is generally non-compact (it is diffeomorphic to $\mathbb{R}$), but from Eq. (\ref{Tl}) we see that once we work in a fixed representation of $H_n$, which we will do in what follows, it effectively becomes $U(1)$.  This makes our formalism loosely resemble Kaluza-Klein theory \cite{KK}---an approach to unification of Maxwell and general relativity theories. There one extends space-time through adding (in a local way) a phase degree of freedom at each space-time point. By introducing a suitable parallel transport on this $5$-dimensional space and postulating an analog of Einstein equations one then recovers coupled gravitational and Maxwell fields. Of course the analogy is only distant, as we are not introducing a parallel transport on our extended phase space, but rather the group structure (cf. Eq. (\ref{prawo})). The motivation for the choice of states is more straightforward---in both in classical and quantum cases characteristic functions possess the same features, provided they are properly defined through Eq. (\ref{fdt}). 

However, the whole $\mathcal{P}_1(H_n)$ turns out to be too large. This happens because a generic $\phi\in\mathcal{P}_1(H_n)$ contains contributions from all possible representations of $H_n$, while we know that only one of them is realized in Nature. To identify the set of physically relevant states within $\mathcal{P}_1(H_n)$ and recover the standard density matrix formalism, we use Gelfand-Naimark-Segal (GNS) construction (see e.g. Ref. \cite{Folland}). Although this is the fundamental tool in algebraic approaches to quantum theory, note that here we are using it in a different manner. In particular, we are not starting from a $C^*$-algebra of observables, but rather from the kinematical group \cite{C*}. Using the GNS construction we can uniquely (up to a unitary transformation) assign to each abstract state $\phi\in\mathcal{P}_1(H_n)$ a triple $(\mathcal{H}_\phi,\pi_\phi,v_\phi)$, where $\pi_\phi$ is a representation of $H_n$ acting in a Hilbert space $\mathcal{H}_\phi$, $v_\phi$ is a normalized cyclic vector, and:
\begin{equation}\label{gns0}
\phi(g)=\langle v_\phi |\pi_\phi(g)\, v_\phi \rangle\,.
\end{equation}
The representation $\pi_\phi$ is generically reducible. It is irreducible if and only if $\phi$ is an extreme point of the state-space $\mathcal{P}_1(H_n)$. Such states will be called ``pure''. Motivated by the commutation relations (\ref{ccr}), we call physical states those abstract states, for which $\pi_\phi$ is a countable multiple of $T^{\hbar}$: $\pi_\phi=\bigoplus_i T^{\hbar}$, $\mathcal{H}_\phi=\bigoplus_i \mathcal{H}_{\hbar}$, since then:
\begin{equation}
\phi(g)=\langle v_\phi |\bigoplus_i T^{\hbar}(g)  v_\phi \rangle = \sum_i\langle v_i |T^{\hbar}(g)  v_i \rangle=\text{tr}[\widetilde{\varrho}_\phi T^{\hbar}(g)],\label{gns}
\end{equation}
where $v_i$'s are the components of $v_\phi$ in each copy of $\mathcal{H}_{\hbar}$, and:
\begin{equation}\label{rho1}
\widetilde{\varrho}_\phi:=\sum_i p_i \big | \frac {v_i}{||v_i||}\big\rangle\big\langle \frac{v_i}{||v_i||}\big |\,,\quad p_i:=||v_i||^2\,,\ \ \sum_i p_i=||v_\phi||^2=1\,.
\end{equation}
Hence, to each physical state $\phi$ we may assign a positive trace-class operator $\widetilde{\varrho}_\phi$ in $\mathcal{H}_{\hbar}$, representing $\phi$. From Eqs. (\ref{Tl}) and (\ref{gns}) we infer that physical states are of a special form:
\begin{equation}\label{phys}
\phi(s,\bm{\eta},\bm{\xi})=\text{e}^{-\text{i}\hbar s}\chi(\bm{\eta},\bm{\xi}),
\end{equation}
where $\chi$ is simply the standard characteristic function (\ref{chi}) of $\widetilde{\varrho}_\phi$. 

The converse also holds, i.e. each abstract state $\phi$ of the form (\ref{phys}) is physical and we can uniquely assign to it a density matrix $\varrho_\phi$. This has been in fact proven in Ref. \cite{Gu}. The operator $\varrho_\phi$ is defined as follows:
\begin{equation}\label{rho2}
\varrho_\phi:=\int_{0}^{\frac{2\pi}{\hbar}}\!\!\!\frac{\text{d}s}{(2\pi)^{2}}\!\int\frac{\text{d}^n \bm{\xi}\, \text{d}^n \bm{\eta}}{(2\pi\hbar)^{n-1}} \, \phi(g)\, T^{\hbar}(g)^{\dagger},
\end{equation}
(provided the above integral exists in the sense of matrix elements, for which it is sufficient that $\chi\in L^1(\mathbb{R}^{2n})$). 
Since $\text{tr}\Big(\text{exp}\Big[\frac{\text{i}}{\hbar}(\eta_j\hat{q_j}-\xi_j\hat{p_j})\Big]\Big)=(2\pi\hbar)^{n}\,\delta^n(\bm{\xi})\delta^n(\bm{\eta})$, we have that:
\begin{equation}
\phi_{\varrho_\phi}(g)=\text{tr}\big[\varrho_\phi T^{\hbar}(g)\big]=\phi(g),\label{reconstr}
\end{equation}
and hence formula (\ref{rho2}) can be viewed as the inverse, with respect to the definition (\ref{fdt}), non-commutative Fourier transform. From Eq. (\ref{reconstr}) and the uniqueness of the GNS construction,
it then follows that $\varrho_\phi$ is the same (up to unitary rotation) as the density matrix $\widetilde{\varrho}_\phi$ from Eq. (\ref{gns}). The representation
(\ref{gns0}) is recovered by spectrally decomposing $\varrho_\phi$ and then going back from Eq. (\ref{gns}) to Eq. (\ref{gns0}). Moreover, if we look from the standard formalism point of view, then we also have $\varrho_{\phi_\varrho}=\varrho$ (since matrix elements of $T^\lambda$ satisfy orthonormality relations like matrix elements of an irreducible representation of a compact group; see Ref. \cite{Gu}). 

Hence, physically relevant states are faithfully represented by functions $\phi\in \mathcal{P}_1(H_n)$ of the form (\ref{phys})\footnote{In fact, for establishing this correspondence we could have used only the formulas (\ref{fdt}) and (\ref{rho2}), but the GNS construction is more general---it can be carried out on an arbitrary locally compact group.}. Please note that restriction to the above form does not break any group symmetry, since the kinematical group $H_n$ is not supposed to act on the abstract state-space $\mathcal{P}_1(H_n)$ as a group of symmetries. This specific form only fixes the representation appearing in the GNS construction to the physically relevant one with $\lambda=\hbar$. From now on we will assume that we work only with the physical states. We have thus achieved the desired reformulation of the standard theory. Density matrices are now secondary objects, constructed from the physical states and the proper representation of the kinematical group through Eq. (\ref{rho2}) (or the GNS construction). We stress that we are dealing here with quantum \emph{statistics} only, as the notion of the linear superposition seems not to be easily visible in the group-theoretical language. Nevertheless the linear structure pertains in our formalism through the GNS construction. It is then an interesting question weather some modification of the group-theoretical framework can be used for construction of a non-linear generalization of quantum mechanics.

The main benefit of the presented reformulation lies, in our eyes, in that it provides a natural transition to the regime $\hbar= 0$, in which one recovers classical statistics. Indeed, if we accept that what is experimentally available are density matrices (for example through the state tomography technique), then we are led to study the irreducible representations of the kinematical group in the case $\hbar= 0$. As can be seen from Eq. (\ref{T0}), the
irreducible representations of $H_n$ become in this regime effectively the irreducible
representations of the Abelian factor-group
$H_n/\{(s,0,0);s\in\mathbb{R}\}=\mathbb{R}^{2n}$, parametrized by $({\bm \eta},{\bm \xi})$. 
Thus the phase degree collapses and the kinematical group turns into the classical phase space. 
Hence, the states that we are naturally led to consider are now functions from $\mathcal{P}_1(\mathbb{R}^{2n})$.
We do not have to worry  about fixing the right, physical representation, like in Eq. (\ref{phys}), as it is already fixed by setting $\hbar=0$ -- these are the representations (\ref{T0}). The crucial point is that Bochner's
Theorem \cite{Folland} states that the functions from $\mathcal{P}_1(\mathbb{R}^{2n})$ are in an
one-to-one correspondence with Borel probability measures on the dual
group $\widehat{\mathbb{R}^{2n}}$, isomorphic to $\mathbb{R}^{2n}$ \cite{Folland}. The duality $\langle\cdot , \cdot\rangle$ is provided by the representation $T^0_{{\bf q},{\bf p}}$ itself:
$\langle ({\bf q},{\bf p}),({\bm \eta},{\bm \xi})\rangle:=T^0_{{\bf q},{\bf p}}({\bm \eta},{\bm \xi})$, $({\bf q},{\bf p})\in \widehat{\mathbb{R}^{2n}}$. Thus, in the classical regime our states, i.e. functions $\phi\in\mathcal{P}_1(\mathbb{R}^{2n})$, can be uniquely identified with Borel
probability measures $\mu_\phi$ on $\widehat{\mathbb{R}^{2n}}\backsimeq \mathbb{R}^{2n}$ and the latter space plays the role of the classical phase space of the system (at least in the context of statistical description). As a result, we recover classical statistical description of the system.

In the case that $\phi\in \mathcal{P}_1(\mathbb{R}^{2n})$ is also in
$L^1(\mathbb{R}^{2n})$, or if we allow for distributions, 
we can explicitly recover $\mu_\phi$ through
the analog of the integral (\ref{rho2}), which now becomes the usual
Fourier transform (we only manually adjust the constant multiplying the measure):
\begin{eqnarray}
& &\text{d}\mu_\phi=\hat{\phi}({\bf q},{\bf p})\,\,\text{d}^n {\bf q} \text{d}^n {\bf p}\,,\label{clas1}\\
& &\hat{\phi}({\bf q},{\bf p}):= \int\frac{\text{d}^n \bm{\xi}\text{d}^n \bm{\eta}}{(2\pi)^{2n}} \, \phi(\bm{\eta},\bm{\xi})\, T^{0}_{{\bf q},{\bf p}}(g)^{\dagger} = \int\frac{\text{d}^n \bm{\xi}\text{d}^n \bm{\eta}}{(2\pi)^{2n}} \, \phi(\bm{\eta},\bm{\xi})\, \text{e}^{-\text{i}(\eta_j q_j-\xi_j p_j)},\label{clas2}
\end{eqnarray}
and $\hat{\phi}$ is a classical probability density in the phase-space.

When applied to composite systems, the above reasoning reveals one interesting aspect of quantum entanglement (see e.g. Ref. \cite{ent} for an introduction into the subject). Namely, when passing to the classical regime, the underlying kinematical group becomes Abelian and the corresponding classical states loose the ability to get entangled, since probability measures on Cartesian products can always be represented as suitable limits of convex combinations of product measures. Hence, the very existence of quantum entanglement may be linked, within our formalism, to the non-Abelian character of the kinematical group. More detailed study of the connection between group-theoretical methods and entanglement is discussed elsewhere \cite{terakurwamy}.

\section{Remarks on observables and dynamics}\label{Obs}
In this Section we briefly remark on the representation of observables and dynamics in our group-theoretical language. We will not be very detailed and mathematically strict here, but rather present a general outline. The easiest observables to deal with are those represented by trace-class operators in the standard language. In our reformulation they are given by complex continuous functions $F$ from $L^1(S_n)$, $S_n:=[0,2\pi/\hbar]\times \mathbb{R}^{2n} \subset H_n$, satisfying:
\begin{equation}\label{obs}
F(g^{-1})=\overline{F(g)}.
\end{equation}
The mean value of $F$ in a state $\phi$ is defined as:
\begin{equation}\label{srednia}
\langle F \rangle_\phi:=\int_{S_n} \text{d}g\, \phi(g) F(g)\,,
\end{equation}
where we have rescaled $dg$ so that it is now equal to $dg=\big[(2\pi)^{2}(2\pi\hbar)^{n-1}\big]^{-1}\text{d}s\,\text{d}^n \bm{\xi}\, \text{d}^n \bm{\eta}$. The integral (\ref{srednia}) is well defined due to the boundedness of $\phi$. To establish the connection with the standard representation of an observable (for the connection to the algebraic approach see remark \cite{C*}), note that to each such $F$ we can assign a hermitian operator $A_F$ by a formula analogous to Eq. (\ref{rho2}):
\begin{equation}\label{pizdalon}
A_F:=\int_{S_n} \text{d}g F(g) T^{\hbar}(g).
\end{equation}
The above integral exists, in the sense of matrix elements, as $F\in L^1(S_n)$. On the other hand, to each trace-class observable $A$ we can assign a continuous function $F_A$ by an analog of Eq. (\ref{fdt}):
\begin{equation}
F_A(g):=\text{tr}[A T^{\hbar}(g)^\dagger].
\end{equation}
Using the same arguments as in the case of density matrices, one can easily show that $F_{A_F}=F$ and $A_{F_A}=A$, thus establishing the correspondence between trace-class observables and functions satisfying Eq. (\ref{obs}).

In the case of observables not representable by trace-class operators, one has to allow for distributions. We will not investigate here which exactly distribution space one needs to consider in order to cover all relevant observables, but only write down the distributions $F_{q_j}$ and $F_{p_j}$ representing the generators $q_j$ and $p_j$:
{\setlength \arraycolsep{1pt}
\begin{eqnarray}
F_{q_j}(s,{\bm \eta},{\bm \xi})&=& \text{i}\hbar\, (2\pi\hbar)^n \text{exp}\Big[\text{i}\hbar s - \frac{\eta_j^2}{4\hbar}\Big]\,\delta^n({\bm\xi})\,\delta(\eta_1)\dots\partial_{\eta_j}\delta(\eta_j)\dots\delta(\eta_n)\\
F_{p_j}(s,{\bm \eta},{\bm \xi})&=& -\text{i}\hbar\, (2\pi\hbar)^n \text{exp}\Big[\text{i}\hbar s - \frac{\xi_j^2}{4\hbar}\Big]\,\delta^n({\bm\eta})\,\delta(\xi_1)\dots\partial_{\xi_j}\delta(\xi_j)\dots\delta(\xi_n). 
\end{eqnarray}}
Higher order polynomials in $q_j$ and $p_j$ are proportional to higher order derivatives of Dirac's delta.

As we have mentioned in the Introduction, the above procedure of recovering standard formalism from the group-theoretical one closely resembles Weyl quantization. Very briefly, Weyl proposed to assign with each Fourier-representable function $F$ on the classical phase space $\Gamma$
\begin{equation}
F({\bf q},{\bf p})=\int\de^n\bm{\xi}\de^n\bm{\eta}\, 
\hat{F}(\bm{\eta},\bm{\xi})\,\text{e}^{\frac{\text{i}}{\hbar}
(\eta_j q_j-\xi_j p_j)}
\end{equation}
an operator
\begin{equation}\label{A_F}
A_F^{Weyl}:=\int\de^n\bm{\xi}\de^n\bm{\eta}
\,\hat{F}(\bm{\eta},\bm{\xi})\,\text{e}^{\frac{\text{i}}{\hbar}
(\eta_j \hat q_j-\xi_j \hat p_j)}
\end{equation}
A comparison of Eq. (\ref{A_F}) with Eqs. (\ref{pizdalon}) and (\ref{rho2}) shows that we basically substitute $\Gamma$ with $H_n$ in the original Weyl formalism. As we have shown in the preceding Section, this substitution leads to some concrete benefits.

Let us move to the representation of dynamics in our language. The dynamical law takes a form of a differential equation imposed on a path $t\mapsto \phi_t$, which should be equivalent to the von Neumann equation for the corresponding density matrix $\varrho_t$. In fact, this equation was derived and analyzed in Ref. \cite{Gu} and we merely quote it here:
\begin{equation}\label{kurwa}
\text{i}\hbar\frac{\partial \phi_t}{\partial t}=\Big[H\Big(\!\!-\text{i}\hbar\partial_{\bm \eta}+\frac{1}{2}{\bm \xi},\,\text{i}\hbar\partial_{\bm \xi}+\frac{1}{2}{\bm \eta}\Big)-H\Big(\!\!-\text{i}\hbar\partial_{\bm \eta}-\frac{1}{2}{\bm \xi},\,\text{i}\hbar\partial_{\bm \xi}-\frac{1}{2}{\bm \eta}\Big)\Big]\phi_t\,, 
\end{equation}
where $H({\bm q},{\bm p})$ is the Hamiltonian, which for simplicity we assume to be of the form $H({\bm q},{\bm p})=T({\bm p})+V({\bm q})$ with $T$ and $V$ analytical. Note that due to the property (\ref{phys}), Eq. (\ref{kurwa}) is, modulo the phase factor, just the quantum Liouville equation \cite{Schleich}, but imposed on the characteristic function $\chi(\bm{\eta},\bm{\xi})$ rather than on the Wigner function $W({\bf q},{\bf p})$.

In the classical limit, as we have argued before, the kinematical group effectively collapses to the classical phase space $\Gamma=\mathbb{R}^{2n}$ and observables become functions (or distributions) on $\Gamma$. The condition (\ref{obs}) now reads: 
\begin{equation}\label{obsclass}
F(-{\bm \eta},-{\bm \xi})=\overline {F({\bm \eta},{\bm \xi})},
\end{equation} 
and from Eqs. (\ref{pizdalon}) and (\ref{T0}) we obtain that:
\begin{equation}\label{pizdalonclass}
A_F({\bm q},{\bm p})= \int\frac{\text{d}^n \bm{\xi}\text{d}^n \bm{\eta}}{(2\pi)^{2n}} \,F({\bm \eta},{\bm \xi})\text{e}^{-\text{i}(\eta_j q_j-\xi_j p_j)}.
\end{equation}
Thus, observables correspond now to real (because of the condition (\ref{obsclass})) functions on the phase-space $\widehat{\mathbb{R}^{2n}}$. Using Eqs. (\ref{clas1}), (\ref{clas2}), and (\ref{pizdalonclass}) the state average (\ref{srednia}) becomes simply the average of $A_F({\bm q},{\bm p})$ with respect to the measure $\mu_\phi$, defined by the state in question $\phi$:
\begin{equation}
\langle F \rangle_\phi=\int\frac{\text{d}^n \bm{\xi}\text{d}^n \bm{\eta}}{(2\pi)^{2n}} \,F({\bm \eta},{\bm \xi})\,\phi({\bm \eta},{\bm \xi})=\int\de\mu_\phi({\bm q},{\bm p})A_F({\bm q},{\bm p}).
\end{equation}
To complete the picture, note that the classical limit of the dynamical law (\ref{kurwa}) for $H({\bm q},{\bm p})={\bm p}^2/2m+V({\bm q})$ reproduces (upon rescaling of ${\bm \eta},{\bm \xi}$ - see the next Section) the classical Liouville equation \cite{Gu}.

\section{Examples of classical limits}\label{Lim}
Here we briefly show with two physical examples how the procedure of moving to classical regime $\hbar= 0$ works in practice. However, let us stress again that it is not the goal of this work to develop another tool for studying classical limits of quantum states, but rather to examine how the formalism of non-commutative characteristic functions leads to the more coherent quantum-classical language and the natural description of classical limit of quantum statistics. Since concrete examples of physically interesting states has been available to us in terms of density matrices $\varrho$ anyway, we have to start from them. From this perspective, our approach obviously brings nothing new to the standard methods of Wigner and Moyal functions, as can it be seen from Eq. (\ref{phys}). Thus, we repeat that the main goal of the present work is conceptual rather than practical.

 The prescription for taking classical limits is rather simple: use the basic formula (\ref{fdt}) to calculate $\phi_\varrho$ for a given matrix $\varrho$ \cite{Gu}. Next, check if there exists, in the distributive sense, a limit $\lim_{\hbar\to 0} \phi_\varrho$ (point limits are too restrictive). If a state $\phi$ is to possess a classical limit at all, we naturally expect that $\big(\lim_{\hbar\to 0} \phi\big)\, \in \mathcal{P}_1(\mathbb{R}^{2n})$, or in other words $\big(\lim_{\hbar\to 0} \phi\big)$ should be a classical characteristic function. If that is the case, we can use the prescription (\ref{clas1}-\ref{clas2}) to retrieve the corresponding probability measure. If not, i.e. $\lim_{\hbar\to 0} \phi$ does not exists, or is not in $\mathcal{P}_1(\mathbb{R}^{2n})$, then the state in question does not possess the classical limit. 

To illustrate the procedure, let us first consider convex mixtures of coherent states, i.e. density operators $\varrho$ with a positive Glauber-Sudarshan $P$-representation \cite{Sudarshan}:
\begin{equation}\label{P}
\varrho=\int_{\mathbb{R}^2} \de\mu(\alpha,\overline{\alpha})|\alpha\rangle\langle\alpha|,
\end{equation}
where $\alpha:=\frac{1}{\sqrt{2\hbar}}(q+\text{i}p)$ and $\mu$ is a probability measure on the classical phase-space $\mathbb{R}^2$. We stress that we consider coherent states here purely kinematicaly, without any explicit or implicit relation to the dynamics. They are defined as the isotropic states minimizing the Heisenberg uncertainty relations, arising from Eq. (\ref{ccr}) and their particular importance for quadratic Hamiltonians does not concern us here. 

Substituting Eq. (\ref{P}) into Eq. (\ref{fdt}) we obtain:
\begin{equation}\label{fP}
\phi_\varrho(s,\eta,\xi)=\int \de\mu(q,p)\,\text{e}^{-\text{i}\hbar s}\,\text{exp}\Big[-\frac{1}{4\hbar}(\xi^2+\eta^2)+\frac{\text{i}}{\hbar}(q\eta-p\xi)\Big].     
\end{equation}
At a first glance, the last term in the integrand in Eq. (\ref{fP}) does not seem to possess any meaningful distributive limit when $\hbar\to 0$. That would be quite counterintuitive, as the matrices of the form (\ref{P}) show a classical-like behaviour: for example the averages of normally ordered observables are equal to the phase-space averages with respect to $\mu$. However, note that the parameters $\eta, \xi$ are just some arbitrary coordinates on the Heisenberg-Weyl group and we are free to re-scale them. Actually, the specific form of the operator $T^\hbar$ in Eq. (\ref{Tl}) was motivated by the physical dimensional analysis (the argument of the exponential function should be physically dimensionless) and in order to recover the group multiplication law (\ref{prawo}) one has to re-scale $\eta, \xi$ by $\hbar$. From another point of view, this rescaling is suggested by the Wigner function (\ref{Wigner}), which can be rewritten as follows:
\begin{equation}
W_\varrho ({\bf q},{\bf p})=\int\frac{\de^n\bm{\xi}\de^n\bm{\eta}}{(2\pi)^{2n}}\text{e}^{-\text{i}(\eta_j q_j-\xi_j p_j)}\chi_\varrho (\hbar\bm{\eta}, \hbar\bm{\xi}).
\end{equation}
If we accept the above arguments, we are led to consider: 
\begin{equation}\label{lim}
\lim_{\hbar\to 0} \phi(s,\hbar\eta,\hbar\xi)
\end{equation}
instead of $\lim_{\hbar\to 0} \phi(s,\eta,\xi)$ as the proper classical limit (compare to the methods of Ref. \cite{Davidovic}). Then from Eq. (\ref{fP}) we obtain that:
\begin{equation}
\phi_\varrho(s,\hbar\eta,\hbar\xi) \xrightarrow [\hbar\to 0]{} \hat{\mu}(\eta, \xi):=\int \de\mu(q,p)\,\text{e}^{\text{i}(q \eta -p \xi)}.
\end{equation}
By Bochner's Theorem (or by an easy direct inspection) $\hat{\mu} \in \mathcal{P}_1(\mathbb{R}^{2})$ and obviously the corresponding probability measure is just the measure $\mu$ itself.

As the next example let us consider the Fock states $|m\rangle$: $\varrho=| m\rangle\langle m|$. Again, we consider them just as kinematical examples. From Eq. (\ref{fdt}) we obtain that:
\begin{equation}
\phi_\varrho(s,\eta,\xi)=\text{e}^{-\text{i}\hbar s} \text{e}^{-\frac{1}{4\hbar}(\xi^2+\eta^2)} L_m\Big[\frac{1}{2\hbar}(\xi^2+\eta^2)\Big],
\end{equation}
where $L_m(x)=m!\sum_{k=0}^m (-1)^k x^k/[(m-k)!(k!)^2]$ is the $m$-th order Laguerre polynomial. Just for illustration's sake, we will considered here a rather uninteresting limit $\hbar \to 0$ of the fixed Fock state $m=\text{const}$. Obviously, this limit does not have much physical sense, but from a {\it purely formal} point of view the vectors $|m\rangle$ are legitimate states in the kinematical space $L^2(\mathbb{R})$ and it is a legitimate question to ask what are their classical limits. Using the prescription (\ref{lim}) we obtain that:
\begin{equation}
\phi_\varrho(s,\hbar\eta,\hbar\xi) \xrightarrow [\hbar\to 0]{} 1\label{chujwyszedl},
\end{equation}
which is trivially a function from $\mathcal{P}_1(\mathbb{R}^{2})$. Hence, after performing the Fourier transform of (\ref{chujwyszedl}), all the matrices $|m\rangle\langle m|$ are mapped in the classical limit to the same probability measure $\delta(q)\delta(p)$. Of course, one would expect that from the form of the energy spectrum of a harmonic oscillator, but as we said before, we consider the limit (\ref{chujwyszedl}) only as a formal exercise. The physically sensible classical limit of the Fock states is given by $\hbar \to 0$, $m \to \infty$, $\hbar m=\text{const}$. In this limit one indeed recovers the classical microcanonical distribution function of the harmonic oscillator, as it was proven in Ref. \cite{Ripamonti} using the closely related method of Wigner functions.

\section{Concluding remarks}\label{Concl}
The next logical step would be to try to apply the developed formalism to systems with compact kinematical groups, like, for example, spin systems with $G=SU(2)$.  The goal would be to describe the well known heuristic prescription: $\hbar\to 0$, $j\to \infty$, $j\hbar=\text{const}$, where $j$ labels the irreducible representations of $SU(2)$, within the presented group-theoretical formalism. One problem immediately arises: the corresponding classical phase-space is a sphere $\mathbb{S}^2$ \cite{Perelomov} and Bochner's Theorem, crucial to our approach, holds only for Abelian groups. Thus, it is not so obvious what mechanism would allow one to recover classical statistics in this case. This is the subject of our further research (for an alternative approach using co-adjoint orbits method see e.g. Ref. \cite{Kus}; for another one based on non-commutative spheres see e.g. Ref. \cite{Madore}). 

Another point is that at this stage our approach lacks a clear operational meaning of the mathematical concepts involved. Perhaps the most operationally flavoured reformulation of quantum theory is the one given by the quantum logic and orthomodular lattices (see Refs. \cite{Mackey, Beltrametti} for an introduction), as it operates directly with the probabilities of outcomes of (idealized) measurements. It also very coherently incorporates classical and quantum statistics within a common language. However, it comes with its own set of problems: the justification for the use of Hilbert spaces for building the lattice and the apparent lack of a clear quantum-to-classical transition mechanism within the formalism. Note that another problem: the justification for the lattice orthomodularity, was solved only very recently by Grinbaum \cite{Grinbaum}, using information-theoretical arguments. 

Summarizing, our work presents an alternative to the standard as well as to the algebraic and lattice approaches to quantum statistics. It incorporates an elegant form of the quantum-to-classical transition. By the latter we mean a clear mechanism of showing how the classical state-space directly arises from the quantum one.  
 
We would like to thank V. Cappellini, J. Derezi\'nski, J. Kijowski, A. Kossakowski, B. Mielnik, J. Naudts, K. $\dot{\text{Z}}$yczkowski and especially M. Ku\'s and M. Przanowski for discussions, and the  Deutsche Forschungsgemeinschaft
(SFB 407, 436 POL), ESF PESC QUDEDIS, EU IP Programme ``SCALA'', and MEC (Spanish Goverment) under contract FIS2005-04627 and Consolider-Ingenio 2010 ``QOIT'' for the financial support.

\end{document}